\documentclass{aa}  

\usepackage{graphicx}
\usepackage{txfonts}
\begin{document}

   \title{Spin-orbit alignment in the very low mass binary regime}

   \subtitle{The L dwarf tight binary 2MASSW J0746425+200032AB}

   \author{L. K. Harding
          \inst{1,2}
          \and
          G. Hallinan
		  \inst{1}
		  \and
		  Q. M. Konopacky
		  \inst{3}\fnmsep\thanks{Dunlap Fellow}
		  \and
		  K. M. Kratter
		  \inst{4}\fnmsep\thanks{Hubble Fellow}
		  \and
		  R. P. Boyle
		  \inst{5}
		  \and
		  R. F. Butler
		  \inst{2}
		  \and
		  A. Golden
		  \inst{6,2}
          }

   \institute{Cahill Center for Astrophysics, California Institute of Technology, 1200 E. California Blvd., MC 249-17, Pasadena, CA 91125, USA\\
              \email{lkh@astro.caltech.edu; gh@astro.caltech.edu}
         \and
             Centre for Astronomy, National University of Ireland, Galway, University Road, Galway, Ireland\\
             \email{ray.butler@nuigalway.ie}
         \and
             Dunlap Institute for Astronomy and Astrophysics, University of Toronto, 50 St. George Street, Toronto M5S 3H4, Ontario, Canada\\
             \email{konopacky@di.utoronto.ca}
         \and
             JILA, University of Colorado, Boulder, CO 80309-0440, USA\\
             \email{kaitlin.kratter@jila.colorado.edu}
         \and
             Vatican Observatory Research Group, Steward Observatory, University of Arizona, Tucson, AZ 85721, USA\\
             \email{boyle@ricci.as.arizona.edu}
         \and
             Department of Genetics (Computational Genetics), Albert Einstein College of Medicine, Bronx NY 10461, USA\\
             \email{aaron.golden@einstein.yu.edu} }

   \date{Accepted April 17, 2013}

  \abstract{Studies of solar-type binaries have found coplanarity between the equatorial and orbital planes of systems with $<$40 AU separation. By comparison, the alignment of the equatorial and orbital axes in the substellar regime, and the associated implications for formation theory, are relatively poorly constrained. Here we present the discovery of the rotation period of 3.32 $\pm$ 0.15 hours from 2MASS J0746+20A - the primary component of a tight (2.7 AU) ultracool dwarf binary system (L0+L1.5). The newly discovered period, together with the established period via radio observations of the other component, and the well constrained orbital parameters and rotational velocity measurements, allow us to infer alignment of the equatorial planes of both components with the orbital plane of the system to within 10 degrees. This result suggests that solar-type binary formation mechanisms may extend down into the brown dwarf mass range, and we consider a number of formation theories that may be applicable in this case. This is the first such observational result in the very low mass binary regime. In addition, the detected period of 3.32 $\pm$ 0.15 hours implies that the reported radio period of 2.07 $\pm$ 0.002 hours is associated with the secondary star, not the primary, as was previously claimed. This in turn refutes the claimed radius of 0.78 $\pm$ 0.1 $R_{J}$ for 2MASS J0746+20A, which we demonstrate to be 0.99 $\pm$ 0.03 $R_{J}$.}


   \maketitle
%

\section{Introduction}

\par Investigation of the spin and orbital properties of solar-type binaries has been underway for several decades \citep[and references therein]{weis74,abt76,bodenheimer78,fekel81,hale94}. A large fraction of close ($<$40 AU) solar-type binaries have been found to have coplanar spins and orbits \citep{hale94}.  Studies by \citet{weis74} yielded similar results over a wider range of spectral classes [B - F]. More recent studies have turned up a diversity of configurations. \citet{jensen04} and \citet{monin06} have reported planar alignment \textit{and} misalignment for wider-separation binaries. Similar to these studies, in the case of very close binaries of semi-major axis $\sim$0.3 AU, there have been some examples where systems exhibit both aligned and misaligned axes \citep{albrecht09,albrecht11}. More recently, \citet{wheelwright11} have also reported coplanarity between HAe/Be binary systems and circumstellar disks. Thus, it appears that while the spin-orbit alignment is common,  there are  exceptions at all orbital separations. 

\par  In this paper, we turn our attention to the spin-orbit alignment of stars at the bottom end of the initial mass function, where binarity is less common \citep{lada06,raghavan10}. In the years following the detection of the first brown dwarf Gl 229B by \citet{nakajima95}, a number of surveys yielded the discovery of a large number of low mass star ($< 0.1$ $M_{\odot}$) and brown dwarf binary systems, e.g. \citet[and references therein]{burgasser07}. Following these discoveries, the introduction of laser guide star (LGS) adaptive optics (AO) systems on ground-based telescopes provided the means of assessing the dynamical mass of such systems \citep{bouy04,dupuy10,konopacky10}. More recently, \citet{konopacky12} obtained resolved LGS AO spectroscopic measurements of individual component rotation velocities for a sample of eleven very low mass (VLM) dwarf binaries (\textbf{$M_{tot} \leq 0.185$ $M_{\odot}$}, as defined by \citet{close03}). These data provided additional parameters for sources  separated by $\sim$1 - 10 AU, but could only be used to tentatively investigate spin-orbit alignment, since other parameters such as individual component rotation periods, and system properties inferred from evolutionary models, such as radii, were still either unknown or poorly constrained.

\begin{table*}
\caption{Properties of 2MASSW J0746425+200032AB}             
\label{table1}      
\centering          
\begin{tabular}{c c c}     
\hline\hline       
Parameter & 2MASS J0746+20A & 2MASS J0746+20B\\ 
\hline                    
	  Rotation period (hrs) & 3.32 $\pm$ 0.15$^{\dag}$ & 2.07 $\pm$ 0.002 \\
           \textit{v} sin \textit{i} (km s$^{-1}$) & 19 $\pm$ 2 & 33 $\pm$ 3 \\
            Equatorial Velocity (km s$^{-1}$) & 36 $\pm$ 4$^{\dag}$ & 56 $\pm$ 2$^{\dag}$ \\
            Period ratio           & 0.62 $\pm^{0.02}_{0.03}$ & ... \\
            \textit{v} sin \textit{i} ratio    & 0.57$\pm^{0.13}_{0.10}$ & ... \\
            Orbital period (yrs) & 12.71 $\pm$ 0.07 & 12.71 $\pm$ 0.07 \\
            Semi-major axis (mas) & 237.3$^{+1.5}_{-0.4}$ & 237.3$^{+1.5}_{-0.4}$ \\
            Inc.$_{ORB}$ (deg)$^{\ddag}$ & 41.8 $\pm$ 0.5 & 41.8 $\pm$ 0.5 \\
            Inc.$_{EQ}$ (deg)$^{\ddag}$	 & 32 $\pm$ 4$^{\dag}$ & 36 $\pm$ 4$^{\dag}$ \\

\hline                    
            Age (log \textit{yrs}) & 9.1 $\pm$ 0.1$^{\dag}$ & 9.1 $\pm$ 0.1$^{\dag}$ \\
            Mass$_{total}$ ($M_{\odot}$) & 0.151 $\pm$ 0.003 & 0.151 $\pm$ 0.003 \\
            Mass ($M_{\odot}$) & 0.078 $\pm$ 0.004$^{\dag}$ & 0.073 $\pm$ 0.004$^{\dag}$ \\
            Lithium? & No & No \\
            Radius ($R_{J}$) & 0.99 $\pm$ 0.03$^{\dag}$ & 0.96 $\pm$ 0.02$^{\dag}$ \\
            Gravity (log \textit{g}) & 5.34 $\pm$ 0.02$^{\dag}$ & 5.34 $\pm$ 0.02$^{\dag}$ \\
            L$_{bol}$ (log $L/L_{\odot}$) & -3.64 $\pm$ 0.02 & -3.77 $\pm$ 0.02 \\
            Abs. mag ($J$) & 11.85 $\pm$ 0.04 & 12.36 $\pm$ 0.10 \\
            Abs. mag ($H$) & 11.13 $\pm$ 0.02 & 11.54 $\pm$ 0.03 \\
            Abs. mag  ($K$) & 10.62 $\pm$ 0.02 & 10.98 $\pm$ 0.02 \\
            References & 1, 3-5 & 1-5\\
\hline                  
\end{tabular}
\tablefoot{$^{\dag}$ Rotation period discovered in this work; all other parameters derived here based on evolutionary models of \citet{chabrier00}.\\$\ddag$Inc.$_{ORB}$ is the system orbital inclination as measured by \citet{konopacky12}, whereas Inc.$_{EQ}$ is the equatorial inclination of each component, calculated in this work.}\\
\tablebib{(1) This work; (2) \citet{berger09}; (3) \citet{konopacky10}; (4) \citet{bouy04}; (5) \citet{chabrier00}.}
\end{table*}

\par Previous studies have shown that the presence of magnetic fields can affect binary formation \citep[and references therein]{mestel77,bodenheimer78,fekel81,li04}, whereby such fields could potentially contribute to a loss of angular momentum on large scales, or the tilting of stellar spins on small scales. Are these concerns of special relevance in the ultracool dwarf regime? M dwarfs later than M3 are now associated with intense magnetic activity, often possessing surface magnetic field strengths of a few kG and greater \citep{reiners07,hallinan08,berger09,west11}. In this paper, we now have  sufficient data to  investigate the orbital properties for the magnetically-active VLM dwarf binary, 2MASS J0746+20AB ($M_{tot}=0.151 \pm 0.003$ $M_{\odot}$). This detection can shed light on the formation of VLM binary stars, and could signal that a scaled-down version of the formation mechanism for solar-type binary systems holds in this regime, despite the presence of a $\sim$1.7 kG magnetic field in this case \citep{antonova08}. Characterizing the fundamental properties of VLM binary star formation is important in establishing a correlation, if any, in the formation and evolution of all types of binary stars.


\section{2MASSW J0746425+200032AB}
\subsection{Properties}

\par 2MASS J0746+20AB is an L dwarf binary (L0+L1.5) that is located at a distance of 12.20 $\pm$ 0.05 pc \citep{dahn02}. It is a tight binary system, with a separation of $\sim$2.7 AU \citep{reid01} and an effective temperature of between 1900 - 2225 K \citep{vrba04}. The latest high-precision dynamical mass measurements yielded a total mass of 0.151 $\pm$ 0.003 $M_{\odot}$ \citep{konopacky10}, initially measured to be 0.146 $\pm^{0.016}_{0.006}$ $M_{\odot}$ by \citet{bouy04}, placing the dwarf in the VLM binary regime. \citet{bouy04} based their individual component masses on a model-derived age estimate. However, \citet{gizis06} questioned these individual mass estimates, and argued against the secondary component being a brown dwarf as predicted by \citet{bouy04}, and instead favored a substellar or low mass star classification.

\par The L dwarf was reported as an active radio source by \citet{antonova08}, who estimated magnetic field strengths of $\sim$1.7 kG based on the detection of a single highly polarized burst of emission. The dominant emission was quiescent, the nature of which is still debated. Following this observation, \citet{berger09} reported radio emission with a rotation period of 2.07 $\pm$ 0.002 hours. Simultaneously, they detected periodic H$\alpha$ emission. The period was the same in both instances, and was consistent with stellar rotation. They also estimated a stellar radius of 0.078 $\pm$ 0.010 R$_{J}$ for 2MASS J0746+20A; in order to do this, they assigned an average \textit{v} sin \textit{i} of 27 $\pm$ 3 km s$^{-1}$ (taken from a number of previous studies \citep{reid02,bailerjones04,reiners08}) to the primary. This assumption was based on the observations of \citet{bouy04}, who established that 70\% of the light contribution was coming from the primary star. Thus, they argued that both the H$\alpha$ and radio emission were also emanating from the primary component of the system. In addition, marginal evidence of periodicity (of a few hours) has previously been reported by \citet{clarke02} and \cite{bailerjones04}. Target properties are shown in Table~\ref{table1}.

\begin{table*}
\caption{Observation details}             
\label{table2}      
\centering          
\begin{tabular}{c c c c c c c c c}     
\hline\hline       
Source &  Date & Length & Exp. & Photometric & Band & Telescope\\
& of obs.& of obs. & time & error & & / instrument\\
& (UT) & ($\sim$hrs) & (s $\times$ coadd) & (\%) & \\ \\
(1) & (2) & (3) & (4) & (5) & (6) & (7)\\ 
\hline                    
	  2MASS J0746+20AB & 2009 Jan 25& 6.0 & 25 $\times$ 1 & 0.21 & I & VATT/4K\\
	  & 2009 Jan 26& 6.8 & 25 $\times$ 1  & 0.28 & I & VATT/4K\\
	  & 2009 Jan 28& 7.4 & 25 $\times$ 1 & 0.24 & I & VATT/4K\\
	  & 2010 Feb 19 & 4.5 & 5 $\times$ 12  & 0.27 & I &VATT/GUFI\\
	  & 2010 Feb 20 & 4.0 & 5 $\times$ 12 & 0.30 & I &  VATT/GUFI\\
	  & 2010 Nov 13 & 4.6 & 5 $\times$ 12 & 0.31 & I & VATT/GUFI\\
	  & 2010 Nov 14 & 5.5 & 5 $\times$ 12 & 0.33 & I & VATT/GUFI\\
	  & 2010 Dec 2 & 6.0 & 5 $\times$ 12 & 0.25 & I &  VATT/GUFI\\
	  & 2010 Dec 12 & 3.0 & 5 $\times$ 12 & 0.29 & I &  VATT/GUFI\\
	  & 2010 Dec 13 & 6.8 & 5 $\times$ 12 & 0.32 & I &  VATT/GUFI\\
	  & 2010 Dec 14 & 7.0 & 5 $\times$ 12 & 0.34 & I & VATT/GUFI\\

\hline                  
\end{tabular}
\tablefoot{Column (1) Target of campaign. (2) Date of each observation in UT. (3) The total time observed per night for each observation, in hours. (4) The exposure time for each observation in seconds, followed by the binning factor used to increase data point S/N. (5) The mean photometric error for light curve data points per night. (6) The VATT Johnson photometric waveband used (I: $\sim$7200 - 9100 \r{A}). (7) Telescope and detector used for a given observation. These observations were taken over 4 separate epochs, spanning $\sim$62 hours of data over a 2 year baseline.}\\
\end{table*}

\subsection{Instrumentation and observations}

\par We obtained observations over a $\sim$2 year baseline for 2MASS J0746+20 to search for periodic photometric variability from either, or both components of the binary system. The observations encompassed 4 separate epochs in January 2009, February 2010, November 2010 and December 2010. These were carried out with the GUFI (Galway Ultra-Fast Imager) and VATT 4K photometers on the VATT telescope\footnote[7]{The Vatican Advanced Technology Telescope (VATT) telescope (1.83 m) facility is operated by the Vatican Observatory, and is part of the Mount Graham International Observatory. Details of VATT detectors can be found here: http://vaticanobservatory.org/VATT/index.php/telescope-instruments/.} using the VATT I-Arizona filter ($\sim$7200 - 9100 \r{A}). GUFI was first commissioned by astronomers at NUI Galway to operate on the 1.5 m Loiano telescope in Bologna, Italy \citep{sheehan08}. It was modified thereafter for use on the VATT telescope, where it is currently stationed as a visitor instrument. GUFI was developed as a dedicated high-speed photometer for a long-term observational campaign that focuses on the periodic variability of a sample of radio detected ultracool dwarfs, and the associated mechanisms responsible for such periodic signals \citep{harding13}. It provides a FOV of $\sim3^{\prime}$ $\times$ 3$^{\prime}$, a plate scale of 0.35$^{\prime\prime}$ pixel$^{-1}$, and performs high-time resolution imaging (e.g. 34 frames per second full frame, with $\sim$2 ms readout rates). Data was taken with exposure times of 5 seconds where frames were later summed in image space to 1 minute to increase the signal to noise (S/N). GUFI was used for the February 2010, November 2010 and December 2010 epochs.

\par We used the VATT 4K photometer for the first epoch of the 2MASS J0746+20 observations in January, 2009, before GUFI's commissioning in May 2009. The VATT 4K is the facility photometer of the VATT observatory, and has a native plate scale of 0.188$^{\prime\prime}$ pixel$^{-1}$ and a FOV of $\sim$12.5$^{\prime}$ $\times$ 12.5$^{\prime}$. Details of observations, and the typical photometric error per given observation are shown in Table~\ref{table2}, where the full sample includes $\sim$62 hours of observations.

\subsection{Data reduction and differential photometry}

\par Standard data reduction techniques were employed via a GUFI pipeline \citep{sheehan08}, which performs bias subtraction (using zero-integration frames) and flat-fielding (using twilight flat-fields). The FOVs of GUFI and the VATT 4K provided up to 20 reference stars for the target field (detector FOV depending). We also carried out photometry for all reference stars in order to measure their level of variability - this comparison ensured that the periodicity was indeed intrinsic to the binary. Reference stars were chosen after assessing their stability (in terms of flux) throughout the night, their isolation on the CCD chip, the properties of their seeing profiles, and their color index with respect to the target. Photometric apertures (in pixels) which provided the highest S/N for the target star were selected for aperture photometry. Since GUFI did not resolve each component of the binary system, the chosen aperture included the combined flux from both stars. Differential photometry was obtained by dividing the target flux by the mean flux of selected reference stars. Finally, after differential photometry was carried out, we assessed the photometric light curves for evidence of any residual systematic trends due to e.g. the effects of increasing/decreasing airmass. These effects were negligible throughout the campaign, where typical seeing at the VATT observatory was $\sim$0.7 - 1.6$^{\prime\prime}$. However, where small trends were observed due to such second order extinction effects, we fitted a linear (or quadratic, trend depending) fit to the data and then applied an inverse form of the chosen fit. Removing trends is important in a variable light curve, since they can introduce aliasing effects in the subsequent spectral analysis (discussed in Section~\ref{variability}). The calculated photometric errors are shown in Table~\ref{table2} (column 5), and the procedure for calculating these errors is outlined in Section~\ref{variability}.

\begin{table}
\caption{Periodic photometric variability from 2MASS J0746+20A}             
\label{table3}      
\centering          
\begin{tabular}{c c c}     
\hline\hline       
Parameter & 2M J0746+20A & 2M J0746+20B\\ 
\hline                    
	   Rotation period (hrs) & 3.32 $\pm$ 0.15 & 2.07 $\pm$ 0.002 \\
            LS$^{\dag}$ period (hrs) & 3.318 & ... \\
            LS sign.$^{\dag}$ ($\sigma$)& $>>$5 & ... \\
            PDM$^{\ddag}$ period (hrs) & 3.32 & ... \\
            PtP$_{target}$ (\%) & 0.40 - 1.52 & ... \\
            Mean $\sigma_{reference}$ (\%) & $\sim$0.36 & ... \\
            Variability & Photometric & Radio/H$\alpha$ \\
            References & 1 & 2 \\
        
\hline                  
\end{tabular}
\tablefoot{$^{\dag}$Period obtained from the Lomb-Scargle (LS) periodogram analysis, as shown in Figure~\ref{fig3} (left). We also include the significance/false alarm probability of the detection as calculated by the LS algorithm.\\$^{\ddag}$Period obtained from the Phase Dispersion Minimization (PDM) technique, shown in Figure~\ref{fig3} (right).}\\
\tablebib{(1) This work; (2) \citet{berger09}.}
\end{table}

\section{Periodic Variability}\label{variability}

\begin{figure*}
   \centering
   \includegraphics[width=17cm]{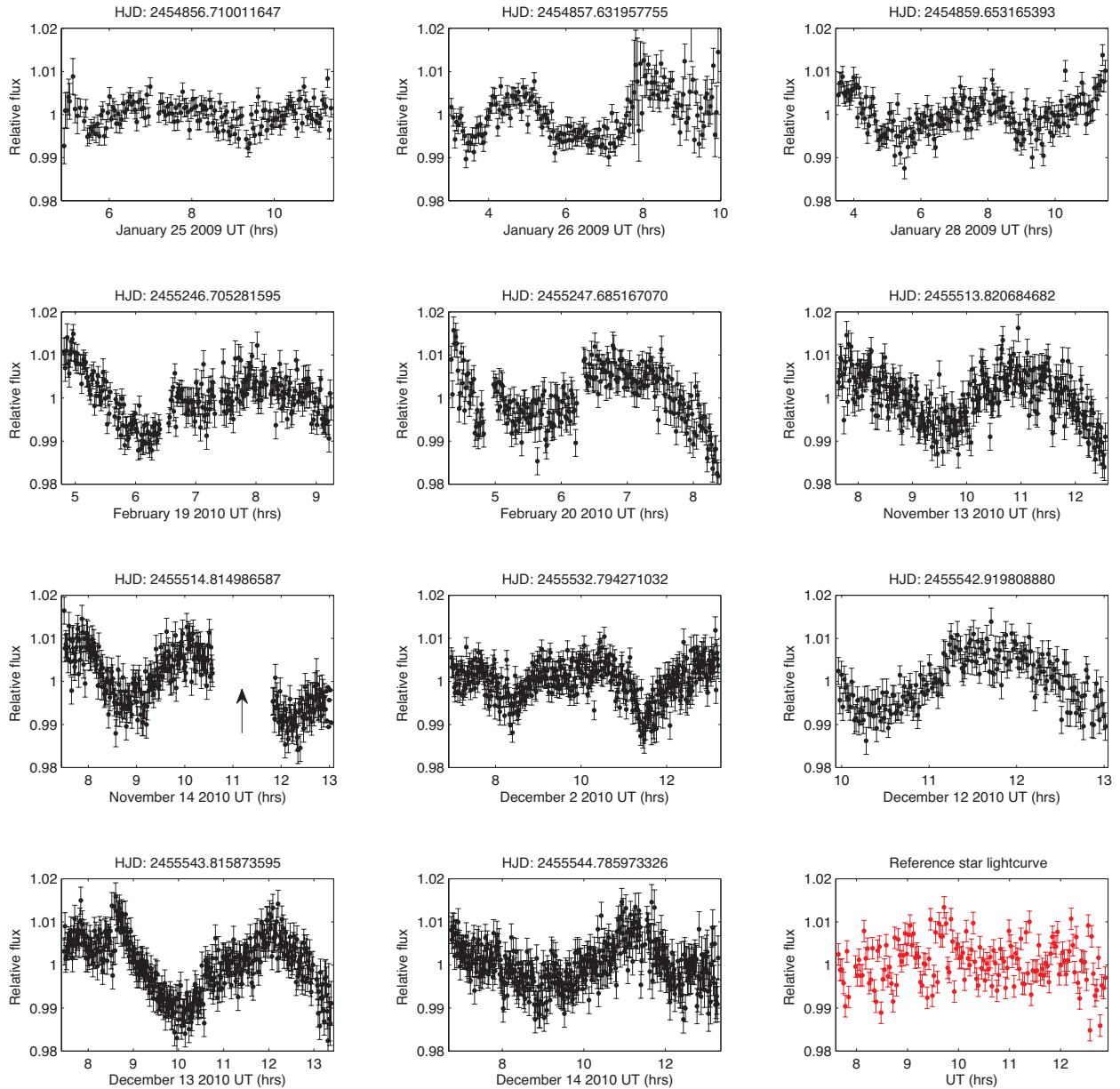}
      \caption{I-band light curves of 2MASS J0746+20AB. Each figure shows relative flux vs. UT times (and dates). We also mark the HJD time above each figure corresponding to the start-point of each observation. All of the data presented here was taken in the `VATT Arizona' I-band broadband filter ($\sim$7200 - 9100 \r{A}), where the baseline extends over $\sim$2 years - as shown in Table~\ref{table1}. \textit{In this work, we report the discovery of consistent periodic variability from 2MASS J0746+20A}, with a period of 3.32 $\pm$ 0.15 hours, and an amplitude variability range of 0.40 - 1.52\%. We note that the full in-depth analysis investigating the stability of the target's phase and amplitude is discussed in other work \citep{harding13}. Some poor weather conditions during constituent epochs were present - e.g. January 25 \& 26 2009 were taken under thin cloud and high wind conditions and thus were binned by a factor of 2 compared to other data sets (Table~\ref{table1}). The vertical arrow marked on the November 14 2010 light curve denotes a period of complete cloud cover, and so data was removed accordingly. The red light curve (bottom right) is an example of a reference star plotted against all others (taken from November 13 2010), illustrating the stability of the chosen reference stars with respect to the target. The mean amplitude variability of the reference stars is shown in Table~\ref{table3}.}
         \label{fig1}
   \end{figure*}
   
\subsection{Variability analysis}

\par The Lomb-Scargle (LS) periodogram \citep{lomb76,scargle82} was used to search for periodic variability in the binary light curves. This technique is very effective for unevenly spaced data, and uses the discrete Fourier Transform (DFT). The algorithms output power spectra with a range of peaks of different amplitudes. The significance of these peaks are then analyzed to a given sigma level as computed by the LS algorithm, which corresponds to possible periodic variability. We selected a range of peaks of $>$5$\sigma$ significance, and then inspected the various solutions by phase folding the light curves to these periods. Comparing these phase folded light curves from both individual nights, and from different epochs, allowed us to assess the level of agreement of phase, for a given period.

\par The standard deviation of the phase fold was also considered. We calculated this via Phase Dispersion Minimization (PDM) techniques \citep{stellingwerf78}. PDM is a least squares fit calculation where the correct period used in the routines produces the least data point scatter in the resulting phase fold. This calculation is called the `PDM Theta statistic' ($\Theta$), where the minimum value is the most likely period solution. We selected a broad range of periods (e.g. 1 - 5 hours), and ran the PDM algorithm giving a corresponding range of $\Theta$ minima. The significance of each $\Theta$ was assessed via 10$^{5}$ Monte Carlo trials, which tests whether any detected $\Theta$ minimum could be a result of noise alone.

\par Finally, we assessed the peak to peak (PtP) amplitude variability of the target light curves (PtP$_{target}$) by fitting a model sinusoidal signal to the raw data using the detected period, and varied the phase and amplitude of the model. A Chi squared ($\chi^{2}$) minimization was performed giving the best fit for a given night's peak to peak amplitude variability. The scatter in the reference star light curves ($\sigma_{reference}$) were established by calculating their standard deviation.

\subsection{Photometric error and period uncertainty estimation}

\par We estimated the error in the relative magnitude of the target star using the \textit{iraf.phot}\footnote[8]{Image Reduction and Analysis. Found here: http://iraf.noao.edu/.} task. The error in magnitude was then converted to an error in flux (since we plot relative flux vs. UT in our photometric light curves). By using \textit{iraf.phot}, we take both formal (e.g. flat-fielding) and informal (e.g. fringing\footnote[9]{Fringing is an additive optical effect at red/NIR wavelengths that occurs in the thinned substrate of back-illuminated CCDs. This is due to atmospheric spectral emission such as OH, and varies in amplitude across the frame. Thus it needs to be removed if varying at a level greater than the amplitude variability of the target star. In the case of 2MASS J0746+20, we assessed this effect and found it to be negligible.}) errors in to account - these are usually quite difficult to assess independently.

\par The period uncertainty was calculated for individual nights using a $\chi^{2}$ test. We could not run this test on the entire baseline at once due to large gaps in the time series. Furthermore, we did not achieve an accurate enough period in consecutive epochs to phase connect the 2 year baseline together. The quoted period and error in this work was therefore derived within epochs, which was found to be consistent for all.

\subsection{Results}

\begin{figure}
   \centering
   \includegraphics[width=9cm]{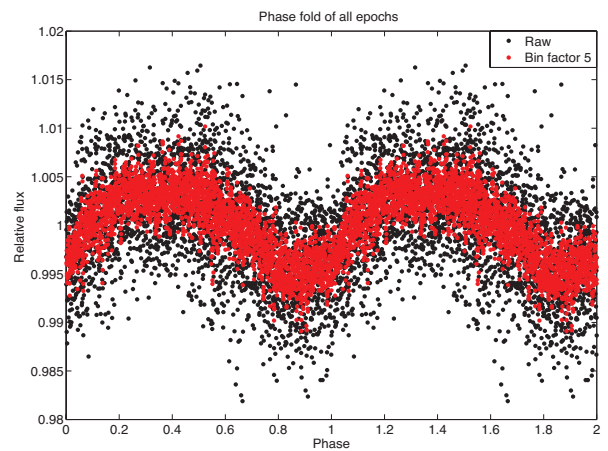}
      \caption{Phase folded light curves of all epochs from the 2MASS J0746+20A observation campaign. The black light curve is folded using the raw light curves as shown in Figure~\ref{fig1}, and the red is the same binned to a factor of 5. Each light curve is phase folded to the detected period of 3.32 $\pm$ 0.15 hours.}
         \label{fig2}
\end{figure}

Although we do not resolve each component of the binary as a point source, most intriguingly, we show optical periodic modulation of 3.32 $\pm$ 0.15 hours in VATT I-band (Figure~\ref{fig1}). We show phase folded light curves for this period in Figure~\ref{fig2}. Therefore, this optical periodic variability originates from the \textit{other} component to that producing the radio emission - reported by \citet{berger09} where the binary exhibited periodic bursts of radio emission of 2.07 $\pm$ 0.002 hours. By adopting the estimated radii in this work (discussed in the next section), in addition to the \textit{v} sin \textit{i} measurements shown in Table~\ref{table1} \citep{konopacky12}, we derive maximum period values of $\sim$4.22 hours and $\sim$2.38 hours for 2MASS J0746+20A and 2MASS J0746+20B, respectively. This infers that we detect the optical rotation period of 2MASS J0746+20A, the primary component of the system. This result is contrary to what was claimed by \citet{berger09}, who attributed the radio period to the primary based upon an incorrect assumption that the \textit{v} sin \textit{i} broadening was due to the primary star, since it is more luminous \citep{bouy04}. 

\begin{figure*}
   \centering
   \includegraphics[width=18cm]{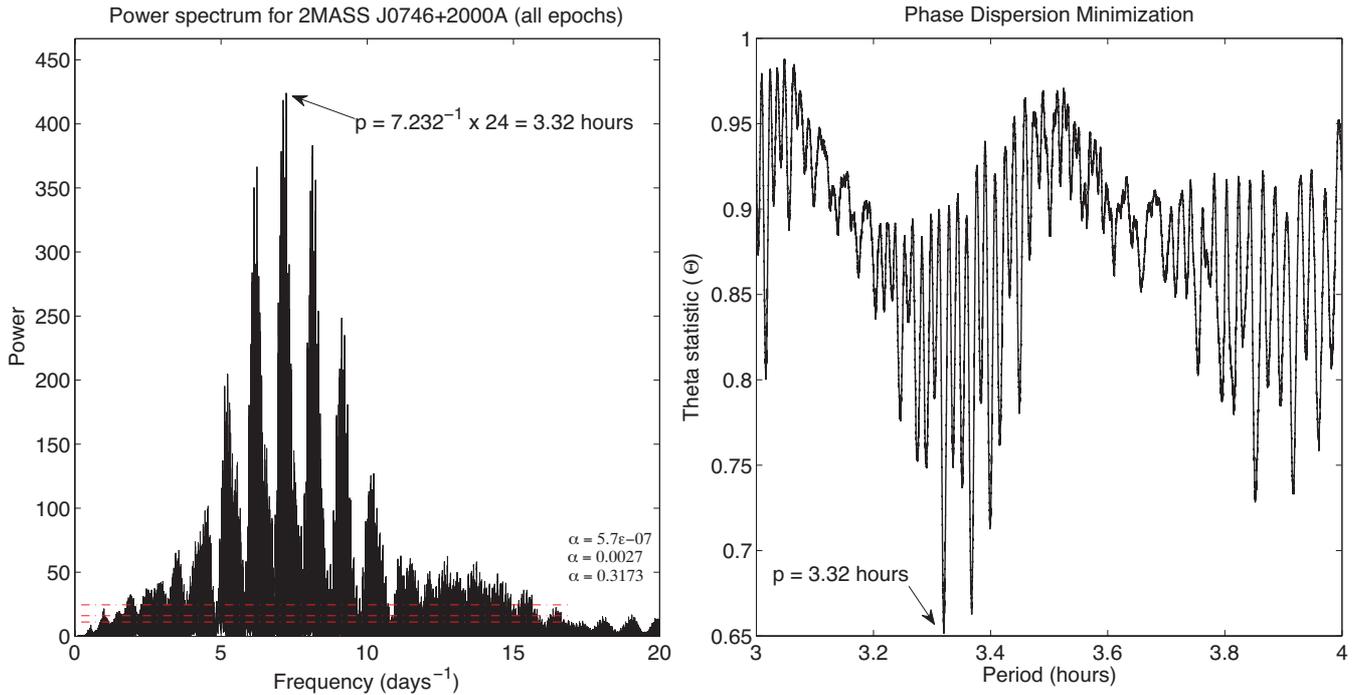}
      \caption{\textit{[LEFT]} Lomb-Scargle periodogram of all 2MASS J0746+20A epochs, calculated from the combined dataset from Figure~\ref{fig1}. The three red dashed-dotted horizontal lines (top-to-bottom) represent a 5$\sigma$, 3$\sigma$ and 1$\sigma$ false-alarm probability of the peaks, as determined by the LS periodogram algorithm. The x-axis is plotted in days$^{-1}$, since each epoch was time-stamped in units of Heliocentric Julian Days (HJD) - for period accuracy over such a lengthy baseline. \textit{[RIGHT]} Phase Dispersion Minimization (PDM) plot of the entire $\sim$2 year observation baseline. Here we plot of the PDM `Theta ($\Theta$) statistic' vs. Period (in hours), where the routines minimize the value at 3.32 hours.}
         \label{fig3}
\end{figure*}

\par A LS periodogram showing the significance of the detection in our work is shown in Figure~\ref{fig3} (left), and the PDM analysis in Figure~\ref{fig3} (right). Both analyses support a period of 3.32 hours for one binary component. The PtP$_{target}$ variability range for all light curves was found to be $\sim$0.40 - 1.52\%, where we also observe changes in light curve morphology throughout the campaign. The source of this behavior is discussed in other work \citep{harding13}. However, in this paper we do investigate the presence of the other binary component - since variable chromospheric emission and an active magnetosphere was reported \citep{berger09}. Thus the presence of magnetic surface features such as hot or cool spots must be considered for both components. We highlight a cluster of reasonably low false alarm probability power spectra around $\sim$12 days$^{-1}$ (or $\sim$2 hours), approximating the radio period. To further assess the possible presence of an underlying period from the secondary component, we fitted a sinusoidal signal with a period of 3.32 hours to each of the raw light curves. We then subtracted this out, and re-ran LS periodograms on each of the remaining data points. For each night, those probabilities were no longer present and thus did not support the presence of the 2.07 $\pm$ 0.002 hour period. We also point out multiple power spectra and $\Theta$ minima around the reported rotation periods in Figure~\ref{fig3} (left and right). These peaks and minima are present as a result of spectral leakage, which is due to large gaps in the data between consecutive observations.

\par Binary system rotation periods are an important parameter in the diagnosis of the system's orbital coplanarity. A radius estimate, in addition to a \textit{v} sin \textit{i} measurement of a system can only loosely infer spin-orbital alignments - the rotation period is an essential variable in the calculation of `\textit{v}', which makes its discovery all the more pertinent. Furthermore, the period reported here has allowed us to infer which period matches each binary member, and thus allowed for an effective estimate of masses and radii, which we discuss in the following section.

\section{An estimate of age, mass and radius}\label{constraint}

\par By adopting the established total system mass of 0.151 $\pm$ 0.003 $M_{\odot}$, the photometric \textit{J} \textit{H} \textit{K} measurements and bolometric luminosity measurements of \citet{konopacky10}, in addition to the lack of detected lithium in the binary dwarf's spectrum \citep{bouy04}, we were able to place constraints on the evolutionary models of \citet{chabrier00}, in determining the mass range and radii of each component. We make no initial assumptions for the age of the system, however young ages were ruled out based on this absence of lithium. Thus we had three measured quantities to estimate the mass track (i.e. \textit{J H K} colors, $L_{bol}$, and Li=0). Once we identified a range of ages that did not contain lithium for either component, we interpolated over a range of masses based on the correlation between the \textit{J H K} colors of \citet{konopacky10}, and those of the \citet{chabrier00} models, and then over the bolometric luminosities - thus establishing the best agreement between each quantity. Most importantly, assuming the stars are coeval, the masses were ultimately constrained since the sum of each component's mass could not be more than the estimated total mass of 0.151 $\pm$ 0.003 $M_{\odot}$ for the system. 

\par We find an age of $\sim$1 - 1.5 Gyr for the binary based on this assessment, as well as individual mass estimates of 0.078 $\pm$ 0.004 M$_{\odot}$ and 0.073 $\pm$ 0.004 M$_{\odot}$ for 2MASS J0746+20A and 2MASS J0746+20B, respectively. These mass estimates are largely consistent with \citet{bouy04}, and in good agreement with \citet{gizis06} and \citet{konopacky10}, and furthermore infer that each component lies at, or just below, the substellar boundary, supporting the prediction of a low mass star classification for the secondary member \citep{gizis06}. The difference in rotational velocity between these stars is most intriguing, considering the similarity of component mass estimates. Perhaps there is a larger difference in mass between each member than what has been inferred from the evolutionary models in the above studies.

\par By contrast, an age of $\sim$1 - 1.5 Gyr identifies the system as a much older binary dwarf than originally predicted by \citet{bouy04}, who found the system to be $\sim$150 - 500 Myr old. This is a large discrepancy. However, we point out that \citet{bouy04} put forward these ages despite the \textit{absence of lithium} in the binary's spectrum - which is expected to be present for stars of this age. We note however, that some studies suggest that lithium can indeed be depleted at younger low mass star ages, due to the effect of episodic accretion \citep{baraffe10}. Although the absence/presence of lithium was originally used as a test for sub-stellarity (e.g. \citet{rebolo92}), it is consequently not necessarily a robust indicator of stellar age. A much younger age is inconsistent with the surface gravity estimates of \citet{schweitzer01}, who compared spectra to the models of \citet{allard01}. Their temperature estimates do agree with those of \citet{bouy04}, but their estimated gravity is too high for a $\sim$150 - 500 Myr old object. These gravity estimates (from high-resolution spectra: log $g$ $\sim$5.0 [K $\mathrm{I}$ $\lambda$7685] - 5.5 [Ca $\mathrm{I}$ $\lambda$5673]; and from low-resolution spectra: log $g$ $\sim$ 6.0) are more consistent with the ages we infer in this work.

\par
\begin{figure}
   \centering
   \includegraphics[width=8.5cm]{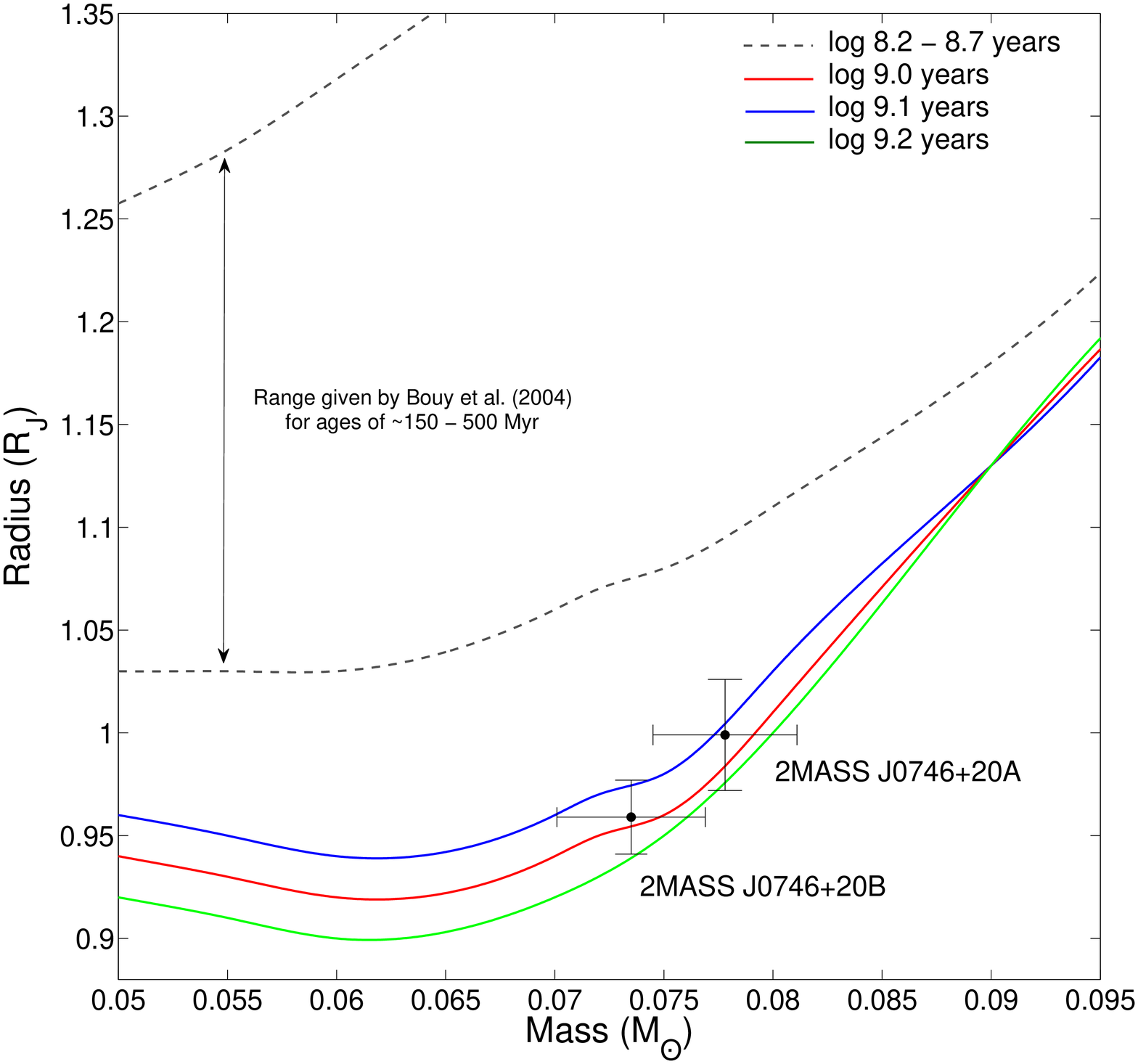}
      \caption{Isochrones of radii ($R_{J}$) vs. mass ($M_{\odot}$) for 2MASS J0746+20AB, for ages of $\sim$1 (log 9.0; red), $\sim$1.25 (log 9.1; blue) and $\sim$1.5 Gyr (log 9.2; green), derived in this work by using the evolutionary models of \citet{chabrier00}. The errors associated with the mass and radii estimates are based on the constraints of the \citet{chabrier00} models for a given age, as outlined in Section~\ref{constraint}. We also include evolutionary tracks for ages of $\sim$150 - 500 Myr (log 8.2 - 8.7 years) as predicted by \citet{bouy04}. These are shown by the grey dash-dotted lines. Based on the work this paper, the binary appears to be a much older age than the range predicted by \citet{bouy04}, which require much larger radii for both components.}
         \label{fig4}
   \end{figure}

 Therefore, adopting the estimates of age and mass above places each star just below 1 $R_{J}$, as shown in Figure~\ref{fig4}, where we estimate radii of 0.99 $\pm$ 0.03 $R_{J}$ for 2MASS J0746+20A and 0.96 $\pm$ 0.02 $R_{J}$ for 2MASS J0746+20B. These predictions are consistent with those of \citet{konopacky10}. However, the radius estimate for 2MASS J0746+20A is inconsistent with the estimate of \citet{berger09}, due to their assumption of a \textit{v} sin \textit{i} of 27 km s$^{-1}$ for 2MASS J0746+20A, and a period of 2.07 $\pm$ 0.002 hours. This rotational velocity is approximately 30\% larger than the established \textit{v} sin \textit{i} of 19 $\pm$ 2 km s$^{-1}$ \citep{konopacky12}, thus incorrectly placing their radius estimate much lower than those predicted by evolutionary models, at 0.78 $\pm$ 0.1 $R_{J}$. Our newly established radii are much more consistent with model predictions, based on our identifying the primary component as the non-radio pulsing star, and strong evidence supporting a spin-orbit alignment to within 10 degrees for both stars. See Table~\ref{table1} for a summary of system properties, including our estimates of surface gravity.

\section{Inferred spin-orbit alignment}

\par
\begin{figure}
   \centering
   \includegraphics[width=9.5cm]{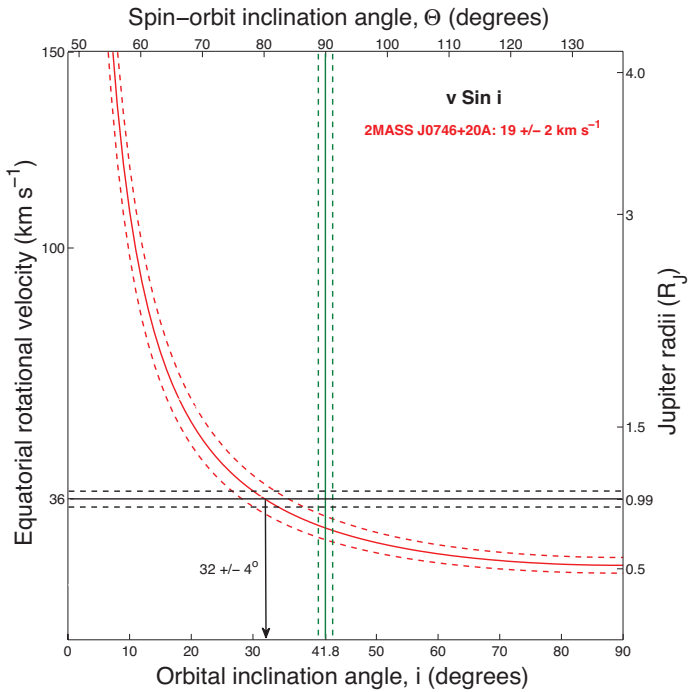}
      \caption{Equatorial rotational velocity (\textit{y-axis, left}) vs. inclination angle of the orbital plane (\textit{x-axis, bottom}). The measured \textit{v} sin \textit{i} of 2MASS J0746+20A is shown by the red solid curve. The dashed lines in all cases represent the associated errors. The green vertical solid line (\textit{x-axis, top}) highlight the alignment of the spin-orbit axes. \textit{Y-axis, right} corresponds to the radius of the dwarf (R$_{J}$ = $\sim$69550 km) as calculated in this work, where we have marked the estimated radius of 0.99 $\pm$ 0.03 R$_{J}$. The black horizontal lines show the corresponding equatorial velocity of 36 $\pm$ 4 km s$^{-1}$. Crucially, this equatorial velocity and the associated alignment only apply for the period of 3.32 $\pm$ 0.15 hours and a radius of 0.99 $\pm$ 0.03 R$_{J}$. The measured equatorial inclination of 32 $\pm$ 4 degrees is marked with the downward arrow, which is within 10 degrees of the orbital inclination angle of 41.8 $\pm$ 0.5 degrees \citep{konopacky12}.}
         \label{fig5}
   \end{figure}

\begin{figure}
   \centering
   \includegraphics[width=9.5cm]{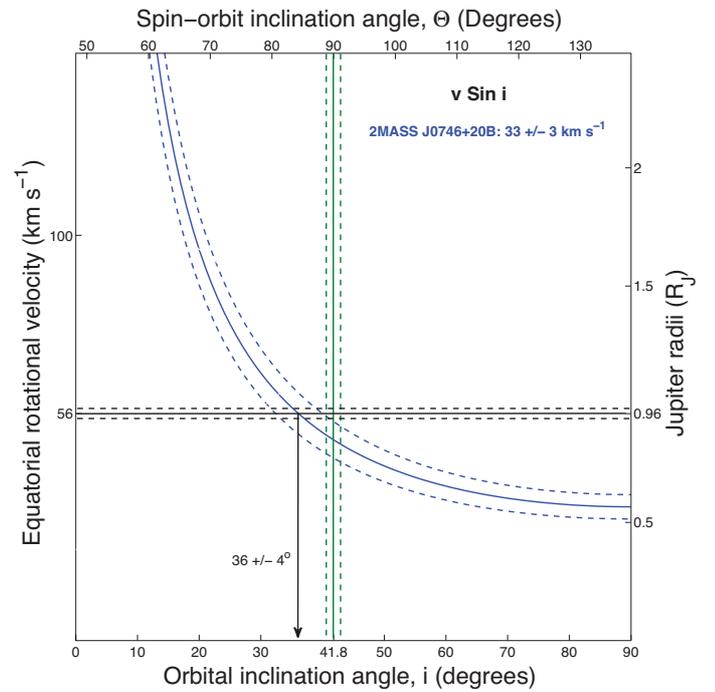}
      \caption{Equatorial rotational velocity (\textit{y-axis, left}) vs. inclination angle of the orbital plane (\textit{x-axis, bottom}) for 2MASS J0746+20B. As before, the measured \textit{v} sin \textit{i} of 2MASS J0746+20B is represented by the blue solid curve, where dash-dotted lines represent the errors in each measurement. We highlight our estimated radius of 0.96 $\pm$ 0.02 R$_{J}$ (\textit{y-axis, right}), which, together with the radio period of 2.07 $\pm$ 0.002 hours \citep{berger09}, infers an equatorial inclination angle of 36 $\pm$ 4 degrees.}
         \label{fig6}
   \end{figure}

 The discovery of the rotation period of 2MASS J0746+20A in this paper, in addition to the spectroscopic observations of \citet{konopacky12} and the radio observations of \citet{berger09}, yield a \textit{v} sin \textit{i} ratio of 0.57$^{+0.13}_{-0.10}$, and a period ratio of 0.62$^{+0.02}_{-0.03}$. By adopting these \textit{v} sin \textit{i} and rotation period measurements, the estimated radii in this work indicate that both components in the binary have their equatorial spin axis aligned with the orbital plane to within 10$^{\circ}$ - consistent with the trends for solar-type binaries with separations $\leq 40$ AU \citep{hale94}. Is this evidence for a common formation pathway across a wide swath of stellar masses? Our best fit indicates that the axes are orientated with respect to the observer at 32 $\pm$ $4^{\circ}$ and 36 $\pm$ $4^{\circ}$, respectively (Figure~\ref{fig5} and Figure~\ref{fig6}). If the measurements are taken at face value, both objects could also be perfectly coplanar, implying that evolutionary models over-predict their radii for an assumed mass and age. The inclination angle of the equatorial spin axis of each component, $sin \; i_{A}$ and $sin \; i_{B}$, were derived as follows:

\begin{equation}
\label{eq1}
      sin \; i_{A} = \frac{v \; sin \; i_{A}}{2 \cdot \pi \cdot r_{A} / P_{A}} \; \; \text{and} \; \; sin \; i_{B} = \frac{v \; sin \; i_{B}}{2 \cdot \pi \cdot r_{B} / P_{B}} \,,
\end{equation}

where \textit{v} is the rotational velocity, $r$ is the radius and $P$ is the period. We estimated the uncertainly in equatorial inclinations above by using Monte Carlo simulations. Each of the variables in Equation~\ref{eq1} had known errors as calculated by  \citet{konopacky12} (\textit{v} sin \textit{i}), this work (r$_{A}$ and r$_{B}$, P$_{A}$) and \citet{berger09} (P$_{B}$), respectively. For each of these variables, we generated 10$^{5}$ copies, where each iteration had random noise (normally distributed) added based on the reported errors. For each of the 10$^{5}$ simulations, we measure the standard deviation of the variables and then calculated a mean value for each. This mean standard deviation over all of the simulations is thus taken as the uncertainty (randomly calculated) in each quantity and applied to the final calculation of the equatorial inclinations in Equation~\ref{eq1} (as shown in Table~\ref{table2} and Figure~\ref{fig5} \& Figure~\ref{fig6}).


\par Finally, we highlight a geometric effect with respect to system spin axes inclinations. The inclinations of the rotation axes of each component to our line of sight may be equal, but this does not necessarily imply that the orbital planes are coplanar. For example, two edge-on spins can still be orthogonal on the sky, and could be coincidentally equal.
   
\section{Discussion: implications for formation}

\par This is the first study to assess the orbital alignment properties of a VLM binary system. The observed orbit-spin alignment is consistent with several different formation pathways. There are numerous binary (and multiple) star system formation theories, the most prominent of which are turbulent core fragmentation, disk fragmentation, and formation via dynamical interactions both during the main accretion phase and after, e.g. \citep{hoyle53,adams89,laughlin94,bonnell94a,tsuribe99,padoan02,bate03,bonnell04,clark08,kratter10,kratter11}.  A nearly coplanar system might result from any of these models, although formation via a dynamical interaction at late times is unlikely; tidal interactions are too weak at these distances, so alignment would be coincidental.

\par Disk fragmentation naturally produces aligned systems, and can also drive components towards equal mass \citep{adams89,laughlin94,bonnell94a,bonnell94b,kratter10}. However, disk fragmentation is more likely for higher mass systems ($>$1 $M_{\odot}$) \citep{kratter08,offner10}. Notably, these models have not been extended down to the brown dwarf regime. \citet{stamatellos09} proposed an alternative disk fragmentation scenario, where binary brown dwarfs are born within the disk of a more massive star. This scenario is  somewhat inconsistent with 2MASS J0746+20AB, as it produces only very high eccentricity binaries (they predict $e>0.6$ compared to this system's $e=0.487$ \citep{konopacky10}). It is also unclear whether ejection would preserve spin alignment.

\par Both core fragmentation and competitive accretion can also produce aligned systems. In the former scenario, fragments may share the core's net angular momentum vector \citep{matsumoto03}. We note that while equal inclinations is a necessary condition for aligned angular momenta, it is not a sufficient one, unless the inclinations are zero. Recent work by \citet{jumper12} has shown that a straightforward extrapolation of the turbulent core fragmentation model to lower masses naturally reproduces the separation distribution of brown dwarfs. However, an analysis by \citet{dupuy11}, finds that the eccentricity distribution of ultracool dwarf binaries is statistically distinct from that of solar type systems, and more consistent with the clustered, competitive accretion model of \citet{bate09,bate12}. Even if such systems are born misaligned, tidal torquing between disks can re-align close systems (see \citet{lubow00}). Interaction with a circumbinary disk, as seen in \citet{bate12}, might also align stars with initially random orientations.

\par As noted above, magnetic interactions, which might be even stronger for fully convective stars, can also alter spin-orbit alignment. The $\alpha\omega$-dynamo, thought to be responsible for the generation of magnetic fields of main sequence stars, no longer functions for fully convective objects. Magnetic fields many orders of magnitude greater in strength than solar-type systems have been shown to exist in the low mass star regime \citep{hallinan07,reiners07,antonova08,berger09}. These fields could have a physical effect on alignment. 2MASS J0746+20AB, at a separation of only $\sim$2.7 AU possesses a large-scale $\sim$1.7 kG magnetic field \citep{antonova08}. Indeed, a strongly misaligned system would be more indicative of dynamical processing. 

\par Although alignment in 2MASS J0746+20AB cannot be used to distinguish between various formation models, more measurements of similar systems might elucidate trends with mass. For example, one might expect the separation at which systems transition from aligned to misaligned to be smaller for lower mass systems, particularly if magnetic fields play a role. Comparing the orbital properties of stars across the mass spectrum may indicate where different formation pathways dominate.

\section{Conclusion}

\par We report on the orbital coplanarity of the L dwarf tight binary 2MASS J0746+20AB. Recently, high-precision dynamical mass and individual rotation velocity measurements were obtained \citep{bouy04,konopacky10,konopacky12}, as well a rotation period for one component \citep{berger09}. We present the discovery of the rotation period from the primary component in this work. From these data, we infer that the binary orbital plane is oriented perpendicular to the stellar spin axes to within 10$^{\circ}$. Such alignment has previously been observed in studies of solar-type binaries \citep{hale94}, and also more massive stars \citep{wheelwright11}. This work is the first direct evidence of spin-orbit alignment in the VLM binary regime. We outline the numerous binary formation models that are consistent with the observed alignment. Further theoretical work, and a larger sample of VLM systems (a campaign that we have already commenced), will place tighter constraints on the most likely formation pathways.

\par The discovery of the 3.32 $\pm$ 0.15 hour period implies that the radio period of 2.07 $\pm$ 0.002 hours is associated with the secondary star and not the primary star, as was suggested in \citet{berger09}. This in turns refutes the claimed radius of 0.78 $\pm$ 0.1 $R_{J}$ for 2MASS J0746+20A. We find that the primary and secondary have radii of 0.99 $\pm$ 0.03 $R_{J}$ and 0.96 $\pm$ 0.02 $R_{J}$ respectively.

\begin{acknowledgements}
This work was largely carried out under the National University of Ireland Traveling Studentship in the Sciences (Physics). LKH gratefully acknowledges the support of the Science Foundation Ireland (Grant Number 07/RFP/PHYF553) as well as Dr. Mark Lang of NUI Galway. We thank the VATT team for their help and guidance. Finally, we would like to thank the referee for their careful reading of our work, and for their valuable input and suggestions on how to improve this manuscript.

\end{acknowledgements}


\begin{thebibliography}{}
\bibitem[Abt \& Levy(1976)]{abt76} Abt, H. A. \& Levy, S. G. 1976, \apjs, 30, 273
\bibitem[Adams et al.(1989)]{adams89} Adams, F. C. \& Ruden, S. P. \& Shu, F. H. 1976, \apj, 347, 959A
\bibitem[Albrecht et al.(2009)]{albrecht09} Albrecht, S. \& Reffert, S. \& Snellen, I. A. G. \& Will, J. N. 2009, \nat, 461, 373
\bibitem[Albrecht et al.(2011)]{albrecht11} Albrecht, S. \& Will, J. N. \& Carter, J. A. \& Snellen, I. A. G. \& de Mooij, E. J. W. 2011, \apj, 726, 68
\bibitem[Allard et al.(2001)]{allard01} Allard, F. \& Hauschildt, P. H. \& Alexander, D. R. \& Tamanai, A. \& Schweitzer, A. 2001, \apj, 556, 357
\bibitem[Antonova et al.(2008)]{antonova08} Antonova, A. \& Doyle, J. G. \& Hallinan, G. \& Bourke, S. \& Golden, A. 2008, \apj, 487, 317
\bibitem[Bailer-Jones(2004)]{bailerjones04} Bailer-Jones, C. A. L. 2004, \aap, 419, 703
\bibitem[Baraffe \& Chabrier(2010)]{baraffe10} Baraffe, I. \& Chabrier, G. 2010, \aap, 521, 44
\bibitem[Bate et al.(2003)]{bate03} Bate, M. R. \& Bonnell, I. A. \& Bromm, V. 2003, \mnras, 339, 577
\bibitem[Bate(2009)]{bate09} Bate, R. B. 2009, \mnras, 392, 590
\bibitem[Bate(2012)]{bate12} Bate, R. B. 2012, \mnras, 419, 3115
\bibitem[Berger et al.(2009)]{berger09} Berger, E. \& Rutledge, R. E. \& Phan-Bao, N. et al. 2009, \apj, 695, 310
\bibitem[Bodenheimer(1978)]{bodenheimer78} Bodenheimer, P. 1978, \apj, 224, 488
\bibitem[Bonnell(1994a)]{bonnell94a} Bonnell, I. A. 1994, \mnras, 269, 837B
\bibitem[Bonnell \& Bate(1994b)]{bonnell94b} Bonnell, I. A. \& Bate, M. R. 1994, \mnras, 271, 999
\bibitem[Bonnell et al.(2004)]{bonnell04} Bonnell, I. A. \& Vine, S. G. \& Bate, M. R. 2004, \mnras, 349, 735
\bibitem[Bouy et al.(2004)]{bouy04} Bouy, H. \& Duch\^ene, G. \& K\"ohler, R. et al. 2004, \aap, 423, 341
\bibitem[Burgasser et al.(2007)]{burgasser07} Burgasser, A. J. \& Reid, I. N. \& Siegler, N. et al. 2007, in Protostars and Planets V, ed. B. Reipurth, D. Jewitt, \& K. Keil (Tucson, AZ: Univ. Arizona Press), 427
\bibitem[Clarke et al.(2002)]{clarke02} Clarke, F. J. \& Oppenheimer, B. R. \& Tinney, C. G. 2002, \mnras, 335, 1158
\bibitem[Clark et al.(2008)]{clark08} Clark, P. C. \& Bonnell, I. A. \& Klessen, R. S. 2008, \mnras, 386, 3
\bibitem[Chabrier et al.(2000)]{chabrier00} Chabrier, G. \& Baraffe, I. \& Allard, F. \& Hauschildt, P. 2000, \apj, 542, 464
\bibitem[Close et al.(2003)]{close03} Close, L. M. \& Siegler, N. \& Freed, M. \& Biller, B. 2003, \apj, 587, 407
\bibitem[Dahn et al.(2002)]{dahn02} Dahn, C. C. \& Harris, H. C. \& Vrba, F. J. et al. 2002, \aj, 124, 1170
\bibitem[Donati et al.(2008)]{donati08} Donati, J. F. \& Morin, J. \& Petit, P. et al. 2008, \mnras, 390, 545
\bibitem[Dupuy et al.(2010)]{dupuy10} Dupuy, T. J. \& Liu, M. C. \& Bowler, B. P. et al. 2010, \apj, 721, 1725
\bibitem[Dupuy \& Liu(2011)]{dupuy11} Dupuy, T. J. \& Liu, M. C. 2011, \apj, 733, 122
\bibitem[Fekel(1981)]{fekel81} Fekel Jr., F. C. 1981, \apj, 246, 879
\bibitem[Gizis \& Reid(2006)]{gizis06} Gizis, J. E. \& Reid, N. I. 2006, \aj, 131, 638
\bibitem[Hale(1994)]{hale94} Hale, A. 1994, \aj, 107, 1
\bibitem[Hallinan et al.(2007)]{hallinan07} Hallinan, G. \& Bourke, S. \& Lane, C. et al. 2007, \apj, 663, L25
\bibitem[Hallinan et al.(2008)]{hallinan08} Hallinan, G. \& Antonova, A. \& Doyle, J. G. et al. 2008, \apj, 684, 644
\bibitem[Harding et al.(2013)]{harding13} Harding, L. K. \& Hallinan, G. \& Boyle, R. P. et al. 2013, \apj, submitted
\bibitem[Hoyle(1953)]{hoyle53} Hoyle, F. 1953, \apj, 118, 513
\bibitem[Jensen et al.(2004)]{jensen04} Jensen, E. L. N. \& Mathieu, R. D. \& Donar, A. X. \& Dullighan, A. 2004, \apj, 600, 789
\bibitem[Jumper \& Fisher(2012)]{jumper12} Jumper, P. H. \& Fisher, R. T. 2012, \apj, submitted (arXiv1206.1045)
\bibitem[Konopacky et al.(2010)]{konopacky10} Konopacky, Q. M. \& Ghez, A. M. \& Barman, T. S. et al. 2010, \apj, 711, 1087
\bibitem[Konopacky et al.(2012)]{konopacky12} Konopacky, Q. M. \& Ghez, A. M. \& Fabrycky, D. C. et al. 2012, \apj, in press
\bibitem[Kratter et al.(2008)]{kratter08} Kratter, K. M. \& Matzner, C. D. \& Krumholz, M. R. 2008, \apj, 681, 375
\bibitem[Kratter et al.(2010)]{kratter10} Kratter, K. M. \& Matzner, C. D. \& Krumholz, M. R. \& Klein, R. I. 2010, \apj, 708, 1585
\bibitem[Kratter(2011)]{kratter11} Kratter, K. M. 2011, in ASP Conf. Ser. 447, Evolution of Compact Binaries, ed. L. Schmidtobreick \& M. R. Schreiber \& C. Tappert, 47
\bibitem[Lada(2006)]{lada06} Lada, C. J. 2006, \apj, 640, L63
\bibitem[Laughlin \& Bodenheimer(1994)]{laughlin94} Laughlin, G. \& Bodenheimer, P. 1994, \apj, 436, 335L
\bibitem[Li et al.(2004)]{li04} Li, P. S. \& Norman, M. L. \& Mac Low, M. M. \& Heitsch, F. 2004, \apj, 605, 800
\bibitem[Lomb(1976)]{lomb76} Lomb, N. R. 1976, \apjs, 39, 447
\bibitem[Lubow \& Ogilvie(2000)]{lubow00} Lubow, S. H. \& Ogilvie, G. I. 2000, \apj, 538, 326
\bibitem[Matsumoto \& Hanawa(2003)]{matsumoto03} Matsumoto, T. \& Hanawa, T. 2003, \apj, 595, 913
\bibitem[Mestel(1977)]{mestel77} Mestel, L. 1977, In IAU Symposium, 75, 213
\bibitem[Monin et al.(2006)]{monin06} Monin, J. \& M{\'{e}}nard, F. \& Peretto, N. 2006, \aap, 446, 201
\bibitem[Nakajima et al.(1995)]{nakajima95} Nakajima, T. \& Oppenheimer, B. R. \& Kulkarni, S. R. et al. 1995, \nat, 378, 463
\bibitem[Offner et al.(2010)]{offner10} Offner, S. S. R. \& Kratter, K. M. \& Matzner, C. D. \& Krumholz, M. R. \& Klein, R. I. 2010, \apj, 725, 1485
\bibitem[Padoan \& Nordlund(2002)]{padoan02} Padoan, P. \& Nordlund, A. 2002, \apj, 576, 870
\bibitem[Raghavan et al.(2010)]{raghavan10} Raghavan, D. \& McAlister, H. A. \& Henry, T. J. et al. 2010, \apjs, 190, 1
\bibitem[Rebolo et al.(1992)]{rebolo92} Rebolo, R. \& Martin, E. L. \& Magazzu, A. 1992, \apj, 389, L83
\bibitem[Reid et al.(2001)]{reid01} Reid, I. N. \& Burgasser, A. J. \& Cruz, K. L. et al. 2001, \aj, 121, 1710
\bibitem[Reid et al.(2002)]{reid02} Reid, I. N. \& Kirkpatrick, J. D. \& Liebert, J. et al. 2002, \aj, 124, 519
\bibitem[Reiners \& Basri(2007)]{reiners07} Reiners, A. \& Basri, G. 2007, \apj, 656, 1121
\bibitem[Reiners \& Basri(2008)]{reiners08} Reiners, A. \& Basri, G. 2008, \apj, 684, 1390
\bibitem[Scargle(1982)]{scargle82} Scargle, J. D. 1982, \apj, 263, 835
\bibitem[Schweitzer et al.(2001)]{schweitzer01} Schweitzer, A. \& Gizis, J. E. \& Hauschildt, P. H. \& Allard, F. \& Reid, I. N. 2001, \apj, 555, 368
\bibitem[Sheehan \& Butler(2008)]{sheehan08} Sheehan, B. \& Butler, R. F., in AIP Conference Proceedings, Volume 984, pp. 162 - 167 (2008)
\bibitem[Stamatellos \& Whitworth(2009)]{stamatellos09} Stamatellos, D. \& Whitworth, A. P. 2009, \mnras, 392 413
\bibitem[Stellingwerf(1978)]{stellingwerf78} Stellingwerf, R. F. 1982, \apj, 224, 953
\bibitem[Tsuribe \& Inutsuka(1999)]{tsuribe99} Tsuribe, T. \& Inutsuka, S. I. 1999, \apj, 526, 307
\bibitem[Vrba et al.(2004)]{vrba04} Vrba, F. J. \& Henden, A. A. \& Luginbuhl, C. B. et al. 2004, \aj, 127, 2948
\bibitem[Weis(1974)]{weis74} Weis, E. W. 1974, \apj, 190, 331
\bibitem[West et al.(2011)]{west11} West, A. A. \& Morgan, D. P. \& Bochanski, J. J. et al. 2011, \aj, 141, 97
\bibitem[Wheelwright et al.(2011)]{wheelwright11} Wheelwright, H. E. \& Vink, J. S. \& Oudmaijer, R. D. \& Drew, J. E. 2011, \aap, 532, A28
\end{thebibliography}
\end{document}